\documentclass[twocolumn,showpacs,aps,prl,letter,amsmath,amssymb,superscriptaddress]{revtex4-1}
\usepackage{bm,color,bbm}
\usepackage{hyperref, mathtools,graphicx,natbib}
\usepackage{graphicx}
\usepackage{amsfonts}    % need for subequations
\usepackage{graphicx}   % need for figures
\usepackage{verbatim}   % useful for program listings
\usepackage{color}      % use if color is used in text
\usepackage{subfigure}  % use for side-by-side figures
\usepackage{hyperref}   % use for hypertext links, including those to external documents and URLs
\usepackage{natbib}
\usepackage{booktabs}
\usepackage{threeparttable}
\usepackage{dcolumn}
\usepackage{multirow}
\usepackage{array}
\usepackage[normalem]{ulem}
\usepackage{soul}

\newcommand{\beq}{\begin{equation}}
\newcommand{\eeq}{\end{equation}}
\newcommand{\bqa}{\begin{eqnarray}}
\newcommand{\eqa}{\end{eqnarray}}

\definecolor{maroon}{rgb}{0.7,0,0}

\definecolor{ngreen}{rgb}{0.3,0.7,0.3}

\definecolor{golden}{rgb}{0.8,0.6,0.1}

\graphicspath{ {./figures/} }

\begin{document}
\title{Observation of the tradeoff between internal quantum nonseparability and external classical correlations }
\author{Jie Zhu}
\affiliation{Key Laboratory of Quantum Information, University of Science and Technology of China, CAS, Hefei, 230026, China, and CAS Center for Excellence in Quantum Information and Quantum Physics, Hefei, 230026, China}
%\author{Meng-Jun Hu}
%\affiliation{Key Laboratory of Quantum Information, University of Science and Technology of China, CAS, Hefei, 230026, China}
%\affiliation{CAS Center for Excellence in Quantum Information and Quantum Physics, Hefei, 230026, China}
%\author{Yue Dai}
%\affiliation{School of Physical Science and Technology, Soochow University, Suzhou, 215006, China}
%\author{Yan-Kui Bai}
%\affiliation{College of Physics Science and Information Engineering and Hebei Advanced Thin Films Laboratory,
%Hebei Normal University, Shijiazhuang, Hebei 050024, China}
\author{Yue Dai}
\affiliation{School of Physical Science and Technology, Ningbo University, Ningbo, 315211, China}
\affiliation{School of Physical Science and Technology, Soochow University, Suzhou, 215006, China}
\author{S. Camalet}
\email{camalet@lptmc.jussieu.fr}
\affiliation{Sorbonne Universit\'{e}, CNRS, Laboratoire de Physique Th\'{e}orique de la Mati\`{e}re Condens\'{e}e,
LPTMC, F-75005 Paris, France}
\author{Cheng-Jie Zhang}
\email{zhangchengjie@nbu.edu.cn}
\affiliation{School of Physical Science and Technology, Ningbo University, Ningbo, 315211, China}
\author{Bi-Heng Liu}
%\email{bhliu@ustc.edu.cn}
\affiliation{Key Laboratory of Quantum Information, University of Science and Technology of China, CAS, Hefei, 230026, China, and CAS Center for Excellence in Quantum Information and Quantum Physics, Hefei, 230026, China}
\author{Chuan-Feng Li}
\affiliation{Key Laboratory of Quantum Information, University of Science and Technology of China, CAS, Hefei, 230026, China, and CAS Center for Excellence in Quantum Information and Quantum Physics, Hefei, 230026, China}
\author{Guang-Can Guo}
\affiliation{Key Laboratory of Quantum Information, University of Science and Technology of China, CAS, Hefei, 230026, China, and CAS Center for Excellence in Quantum Information and Quantum Physics, Hefei, 230026, China}
\author{Yong-Sheng Zhang}
\email{yshzhang@ustc.edu.cn}
\affiliation{Key Laboratory of Quantum Information, University of Science and Technology of China, CAS, Hefei, 230026, China, and CAS Center for Excellence in Quantum Information and Quantum Physics, Hefei, 230026, China}

% USE FOR BOTH
\begin{abstract}
\noindent The monogamy relations of entanglement are highly significant. However, they involve only amounts of entanglement shared by different subsystems. Results on monogamy relations between entanglement and other kinds of correlations, and particularly classical correlations, are very scarce. Here we experimentally observe a tradeoff relation between internal quantum nonseparability and external total correlations in a photonic system and found that even purely classical external correlations have a detrimental effect on internal nonseparability. The nonseparability we consider, measured 
by the concurrence, is between different degrees of freedom within the same photon, and the external classical correlations, measured by the standard quantum mutual information, are generated between the photons of 
a photon pair using the time-bin method. 
Our observations show that to preserve the internal entanglement in a system, it is necessary to maintain low external correlations, including classical ones, between the system and 
its environment.
\end{abstract}

\date{\today}% It is always \today, today,

%USE FOR REVTEX
%\pacs{03.65.Ta}
\maketitle

%%%%%%%%%%%%%%%%%%%%%%%%%%%%%%%%%%%%%%%%%%%%%%%%%%%%%%%%%%%%%%%%%%%%%%%%%%%%%%%%%%%%%%%%%%%%%%%%%%%%%%%%%%%%%%%%%%%%%%%%%%

\noindent{\bf INTRODUCTION}

\noindent Since the Einstein-Podolsky-Rosen (EPR) paradox \cite{epr} was first proposed in 1935, quantum entanglement \cite{HHHH} has drawn considerable attention. Subsequently, the Bell inequality \cite{bellinequality} provided a further demonstration that quantum entanglement differs from any classical correlation. As part of the ongoing progress in the research on quantum entanglement, the properties of multipartite quantum systems must be characterized. Most explorations of multipartite systems to date have focused on quantum entanglement. For example, the important traditional entanglement monogamy relation for three qubits \cite{CKWmonogamy,gilad2018}, 
say $A$, $B$ and $C$, states that the entanglement shared by qubits $A$ and $B$ and that shared by qubits $A$ and $C$ limit each other. This monogamy relation has been generalized to the multipartite systems case \cite{CKWmonogamy2}. Recently, Camalet derived a new type of monogamy relation that is a tradeoff between the internal entanglement within one system and the external entanglement of that system with another system \cite{C3,C4}, which has been observed experimentally \cite{ZHDBCZLGZ}. 

\begin{figure}
\includegraphics[scale=0.33]{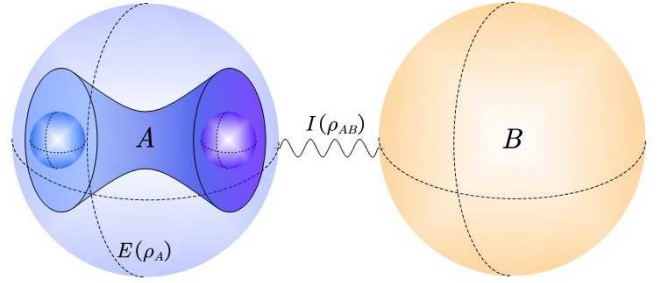}
\caption{\textbf{Theoretical sketch.} 
The internal quantum entanglement (or nonseparability) of system $A$, quantified by $E(\rho_{A})$, and the total correlations between $A$ and another system $B$, quantified by $I(\rho_{AB})$, limit each other. Quantum entanglement (nonseparability) can exist between different subsystems (different degrees of freedom) of $A$. In addition, system $A$ and system $B$ can also be viewed as a system ($A$) and its environment ($B$), respectively. Therefore, reduction of the correlations (including classical correlations) between the system and its environment can be highly beneficial for preserving the internal entanglement of the system.}\label{theory}
\end{figure}
    
On the other hand, relations between quantum entanglement and other kinds of correlations, and particularly classical correlations, have 
attracted far less attention.
It is well known that, while the distribution of quantum correlations, such as entanglement, is constrained, classical correlations can be shared freely. The question of whether any limitation exists between quantum correlations 
(e.g. entanglement or nonseparability) and classical correlations then naturally arises. This issue has been addressed for the three-system scenario mentioned above \cite{KW} and, more recently, 
for a composite system correlated to another system \cite{C}. In this last case, 
it has been demonstrated that internal entanglement has 
a tradeoff relation not only 
with external entanglement but also with other forms of 
external correlations, total correlations for instance, as 
illustrated in Fig. \ref{theory}. 
In particular, even purely classical external correlations limit internal entanglement and vice versa.

In this study, we provide an experimental demonstration of this monogamy relation in a photonic system. Instead of internal entanglement, we consider the analogue for two 
degrees of freedom of a single system, a photon in our experiments, known as quantum nonseparability 
\cite{ZHDBCZLGZ,schumacher1991,hasan2020}.
The time-bin method \cite{timebinfirst,francesco2018} is used to prepare purely classical correlations between two photons. 
This is a novel using of this technique for that purpose.
The considered tradeoff relation 
%is vital in the investigation of quantum network for quantum communication, quantum key distribution, even in quantum computation. It is also 
may play an important role in the evolution of open systems \cite{ines2017,simon2020}, where the internal entanglement (or nonseparability) of the open system 
and the external correlations with the system environment, 
influence each other continuously. 
Therefore, our results show that it is necessary to maintain low external correlations, including classical correlations, to allow more of the internal entanglement or nonseparability of the system to be preserved.

%%%%%%%%%%%%%%%%%%%%%%%%%%%%%%%%%%%%%%%%%%%%%%%%%%%%%%%%%%%%%%%%%%%%%%%%%%%%%%%%%%%%%%%%%%%%%%%%%%%%%%%%%%%%%%%%%%%%%%%%%%

\hfill

\noindent{\bf THEORETICAL BACKGROUND}

\noindent Consider any finite-dimensional system $A$, which
consists of two subsystems (or has two degrees of freedom), 
and any other system $B$, see Fig.\ref{theory}. 
The entanglement (or the quantum nonseparability) between the two subsystems (or degrees of freedom) of $A$ and the total correlations between $A$ and $B$ limit each other \cite{C}. This relation can be described quantitatively using an inequality that involves an entanglement monotone $E$ and a correlation monotone $C$. Monotone $C$ is required to vanish for product states and to be non-increasing under local operations, i.e., operations that do not affect either $A$ or $B$ \cite{C,C2}. Monotone $E$ must vanish for separable states and be non-increasing under local operations 
and classical communication \cite{V,HHHH}. 

The inequality described above can be written more specifically as
\begin{equation}
E(\rho_A)  \le \xi [C(\rho)]  , \label{mineq}
\end{equation}
where $\rho$ is the global state of $A$ and $B$, $\rho_A$ is the state of $A$, and $\xi$ is a non-increasing function. For any number of correlations $c$, there are states $\rho$ such that $C(\rho)=c$ and the two sides of the inequality are infinitely close to each other. 
We note that $C$ in Eq. \eqref{mineq} can be an entanglement monotone because such a measure is also a correlation monotone. This inequality thus generalizes the relation between the internal and external entanglements studied in Refs. \cite{C3,ZHDBCZLGZ}. 
Equation \eqref{mineq} has been obtained by assuming that $E$ is convex and that $C$ satisfies the following requirement.
When $A$ and $B$ are in a pure state $|\psi\rangle$, 
$C(|\psi\rangle \langle \psi|)$ is only dependent on the nonzero 
eigenvalues $\lambda_1 (\rho_A), \ldots$ of $\rho_A$, i.e., 
$C(|\psi\rangle \langle \psi|)=f(\lambda_1 (\rho_A), \ldots)$.
Equation \eqref{mineq} can be derived when the function $f$ is continuous. 

\begin{figure}
\includegraphics[scale =0.4]{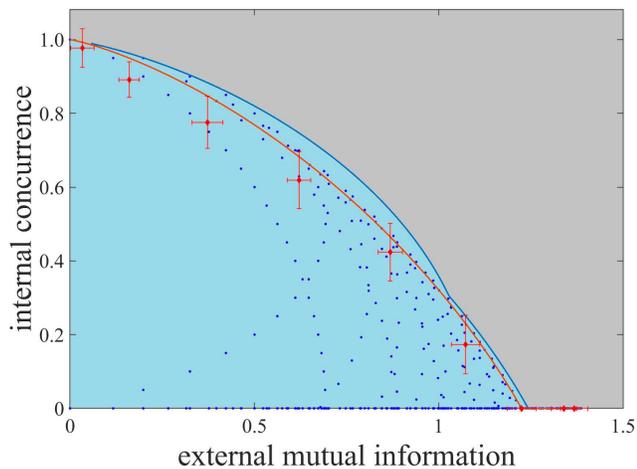}
\caption{\textbf{Relationship between the internal quantum nonseparability and the external classical correlations.} The zero axes and the blue line delimit the allowed region when 
systems $A$ and $B$ are not entangled, which is shown as the light blue shaded region. The grey region cannot be reached with separable states. 
When the mutual information between the non-entangled $A$ and $B$ is larger than $\ln(2\sqrt{3})$, the internal concurrence of $A$ necessarily vanishes. The red line is obtained using Eqs. \eqref{Iexp} and \eqref{Eexp}. The blue dots represent 
theoretical sampling results for the two-parameter family 
of states that generalizes Eq. (\ref{expst}), see Supplementary 
Materials \cite{SM}, and the red dots with the error bars represent the experimental results. The error bars result from the imperfections of the interferometer and the deviations of the polarization devices.  } 
\label{theoreticalexper}
\end{figure}

Because inequality \eqref{mineq} holds when $C$ is a measure of the total correlations, it implies that even purely classical correlations between $A$ and $B$ have a detrimental effect on the internal entanglement of $A$. This can be seen clearly in the case where systems $A$ and $B$ are in a classical-classical state
\begin{equation}
\rho=\sum_{i,j} p_{ij} 
|i\rangle \langle i| \otimes|j\rangle \langle j| ,
\label{ccs}
\end{equation}
where $\{|i\rangle\}$ (\{$|j\rangle$\}) denote 
orthonormal states of $A$ ($B$) and $ p_{ij}$ are probabilities that sum to unity. For such a state $\rho$, not only there is no entanglement between $A$ and $B$, but also the quantum discord measures vanish \cite{MBCPV}. 
The classical-classical states \eqref{ccs} obey Eq. \eqref{mineq} with $\xi$ being replaced with a non-increasing function $\zeta$, which is lower than $\xi$, see Fig. \ref{theoreticalexper} \cite{C}. 

The total correlations are usually quantified using the mutual information 
\begin{equation}
I(\rho)=S(\rho_A)+S(\rho_B)-S(\rho), \label{mi}
\end{equation}
where $S$ is the von Neumann entropy, which is readily computable for any global state $\rho$. For general states,  
the mutual information is not larger than $2\ln d$, where $d$ is the Hilbert space dimension of $A$. For the classical-classical states \eqref{ccs}, the mutual information cannot exceed $\ln d$. This measure is a correlation monotone \cite{MPSVW}. We use it in the following to evaluate the correlations between $A$ and $B$. In this case, one has $\zeta(\ln d)=0$ for any $E$. In other words, for a classical-classical state $\rho$, the internal entanglement 
$E(\rho_A)$ must necessarily vanish when $I(\rho)$ reaches the corresponding 
maximum value of $\ln d$. Note that here inequality \eqref{mineq} with $C=I$ and $\zeta$ in place of $\xi$ is actually valid for all separable states $\rho$. 

For a system $A$ that consists of two two-level subsystems (or two two-level degrees of freedom), the concurrence is a familiar entanglement monotone that can be evaluated for any state $\rho_A$ \cite{W}. Because it is a convex roof measure, it is convex \cite{V}. Its maximum value is $1$. When $E$ is the concurrence and $C$ is the mutual information \eqref{mi}, 
the function $\zeta$ can be obtained using the results of 
Ref. \cite{VAD}; see the Supplementary Materials \cite{SM}. 
This function vanishes in the interval $[\ln (2\sqrt{3}),2\ln 2]$. 
In the interval $[0,\ln (2\sqrt{3})]$, 
its inverse function is given by
\begin{eqnarray}
&&\zeta^{-1}(e)=\max\Big\{  \mu(1+e)+\mu(1-e)+(1-e)\ln (3)/2,\nonumber\\
&&\ \ \mu(1+e-\kappa(e))+\mu(1-e-\kappa(e))
-\kappa(e)\ln \kappa(e)  \Big\}  ,
\label{IC}
\end{eqnarray}
where $\mu(e)=-e \ln(e/2)/2$ and $\kappa(e)=(\sqrt{4-3e^2}-1)/3$. 
The function $\zeta$ is shown as a blue line in Fig.\ref{theoreticalexper}.

We now consider a system $A$ that consists of 
two two-level subsystems (or two two-level degrees of freedom)
and a four-level system $B$ in a classical-classical state of the form
\begin{multline}
\rho^{(p)}=p^2 |00\rangle \langle 00| \otimes 
|\alpha \rangle \langle \alpha|
+(1-p)^2 |+ \rangle \langle + | \otimes |\beta \rangle \langle \beta|\\
+p(1-p) \Big(| 11 \rangle \langle 11 | \otimes 
|\gamma \rangle \langle \gamma|
+ |- \rangle \langle - | \otimes |\delta \rangle \langle \delta| \Big) , 
\label{expst}
\end{multline}
where $p \in [0,1]$, $|\alpha \rangle$, $|\beta \rangle$, 
$|\gamma \rangle$, and $|\delta \rangle$ are orthonormal states, and 
$| 00 \rangle=| 0 \rangle \otimes | 0 \rangle$, 
$| 11 \rangle=| 1 \rangle \otimes | 1 \rangle$, and 
$|\pm\rangle
=(|0 \rangle \otimes | 1\rangle\pm|1 \rangle \otimes |0\rangle)/\sqrt{2}$ 
with orthonormal states $|0\rangle$ 
and $|1 \rangle$ of the subsystems 
of $A$. Below, we will see how states \eqref{expst} can be prepared experimentally. The corresponding mutual information \eqref{mi} is given by 
\begin{equation}
I\big(\rho^{(p)}\big)=-2p\ln p - 2(1-p)\ln (1-p) , 
\label{Iexp}
\end{equation}
which reaches all values in the interval $[0,2\ln 2]$ when $p$ varies from $0$ to $1/2$. The concurrence of the reduced density operator $\rho_A^{(p)}$ is given by 
\begin{equation}
E\big(\rho_A^{(p)}\big)
= \max \Big\{ 0, (1-2p)(1-p)-2\sqrt{p^3(1-p)} \Big\} ,
\label{Eexp}
\end{equation}
which decreases from $1$ to $0$ as $p$ increases and vanishes when $p > 0.302$; see the Supplementary Materials \cite{SM}. 

The curve described by $(I\big(\rho^{(p)}\big),E\big(\rho_A^{(p)}\big))$, as $p$ 
varies from $0$ to $1/2$, is shown as a red line in Fig. \ref{theoreticalexper}. The extreme cases where $p=0$ and $p=1$ can be 
readily understood from expression \eqref{expst}. 
For the global state 
$\rho^{(0)}
=|+ \rangle \langle + | \otimes |\beta \rangle \langle \beta|$, 
$A$ and $B$ are uncorrelated and the two subsystems of $A$ are maximally entangled, which means that $I=0$ and $E=1$. For 
$\rho^{(1)}
=| 00 \rangle \langle 00| \otimes |\alpha \rangle \langle \alpha|$, 
$A$ and $B$ are uncorrelated and the two subsystems of $A$ are also uncorrelated, which means that $I=E=0$. For $p=1/2$, $\rho_A^{(1/2)}$ is the maximally mixed state of $A$ and thus the two subsystems of $A$ are uncorrelated and $E=0$. For any $p$, the mutual information \eqref{Iexp} and the concurrence \eqref{Eexp} satisfy inequality \eqref{mineq} with the function $\zeta$ given by Eq. \eqref{IC}, i.e., in Fig. \ref{theoreticalexper}, the red line is entirely in the light blue shaded region. 
In the Supplementary Materials, a two-parameter family of 
states that generalizes Eq. \eqref{expst} is considered. The blue 
dots in Fig. \ref{theoreticalexper} correspond to such states.

\begin{figure*}
\includegraphics[scale=0.32]{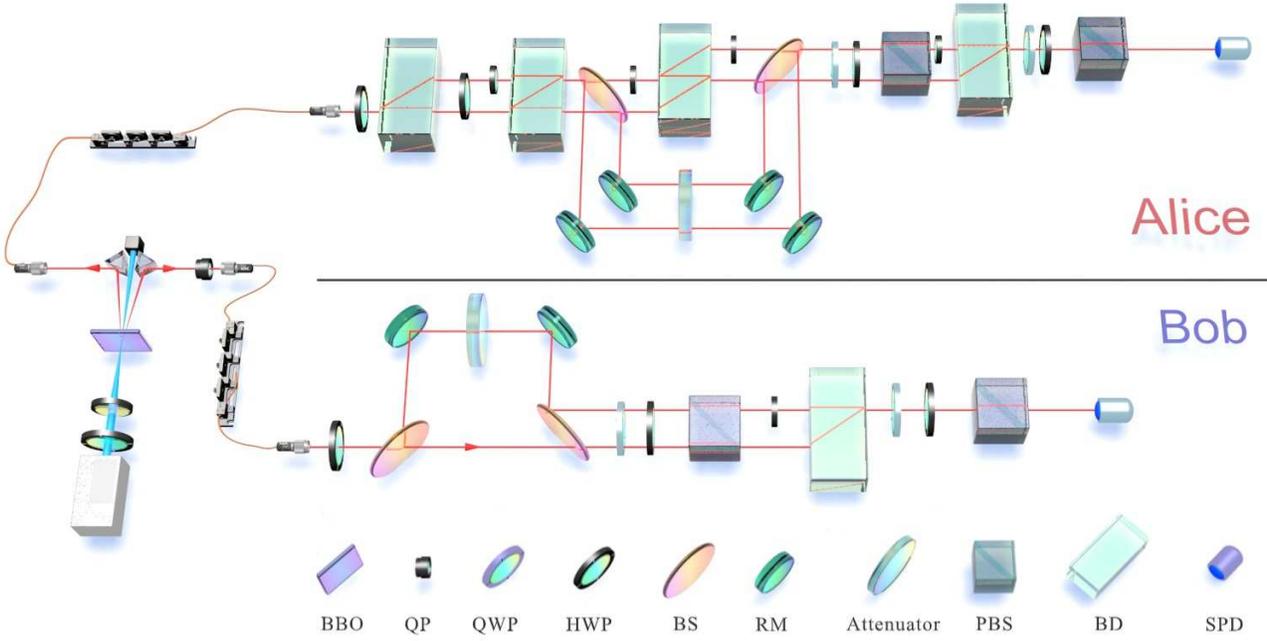}
\caption{\textbf{Experimental setup.} The type-I-phase-matching spontaneous parametric down-conversion process generates an entangled photon pair that is decohered after passing through a sufficiently long quartz plate. The two correlated photons are then distributed to Alice (the upper part) and Bob (the lower part) through optical fibers. In Alice's section, the polarization and the path are nonseparable. BBO: $\beta$-barium-borate crystal; QP: quartz plate; QWP: quarter-wave plate; HWP: half-wave plate; BS: beam splitter; RM: reflective mirror; PBS: polarizing beam splitter; BD: beam displacer; SPD: single photon detector.}\label{setup}
\end{figure*}

%%%%%%%%%%%%%%%%%%%%%%%%%%%%%%%%%%%%%%%%%%%%%%%%%%%%%%%%%%%%%%%%%%%%%%%%%%%%%%%%%%%%%%%%%%%%%%%%%%%%%%%%%%%%%%%%%%%%%%%%%%

\hfill

\noindent{\bf EXPERIMENT}

\noindent To test the relation between 
internal quantum nonseparability and external correlations experimentally, we first prepare some states as per the form of Eq. \eqref{expst} and measure them to evaluate the corresponding concurrence and mutual information. Polarization and path degrees of freedom are used for the preparation of these states. As Fig. \ref{setup} shows, pairs of entangled photons are generated, and the two photons of a pair travel along two different paths via single-mode fibers. The upper path belongs to Alice, which receives photon $A$, and the lower path belongs to Bob, which receives photon $B$. In the following, we will describe these two parts of the setup in detail.

In the entangled-photon-pair source, the entangled state $|\psi\rangle = \cos\theta|HH\rangle
+\sin\theta|VV\rangle$, where $|H\rangle$ 
and $|V\rangle$ denote, respectively, horizontal and vertical polarization states of a photon, is produced via a type I phase-matching spontaneous parametric down-conversion (SPDC) process in a joint $\beta$-barium-borate (BBO) crystal \cite{P.G.Kwiat1999}. The angle $\theta$ is 
adjustable.
The laser used here is a semiconductor laser with a wavelength of 404 nm and power of approximately 100 mW. Then, a suitably long quartz plate causes this state to be completely decohered into 
the mixed state  $\rho_d = \cos^2\theta|HH\rangle \langle HH| + \sin^2\theta|VV\rangle\langle VV|$. 
The two photons are subsequently distributed to the
Alice's and Bob's parts of the setup, 
where operators act on them to produce the target states as shown in Eq. (\ref{expst}). In both Alice's and Bob's parts, there are 
an up-path and a down-path.
In the following, we use two conventions for both 
photons A and B: (\textrm{i}): both $|H\rangle$ and the path state $|\text{up-path}\rangle$ represent $|0\rangle$; and (\textrm{ii}): both $|V\rangle$ and the path state $|\text{down-path}\rangle$ represent $|1\rangle$.

In Alice's part, two operations are performed on photon A with different probabilities. The first operation is $U_1: (|0\rangle, |1\rangle) \rightarrow (|11\rangle, |+\rangle)$, which has probability $1-p$. The second operation is 
$U_2: (|0\rangle, |1\rangle) \rightarrow (|00\rangle, |-\rangle)$, which has probability 
$p$. We fix the two half-wave plates (HWPs) after the first beam displacer (BD) at angles of $45^\circ$ and $22.5^\circ$, respectively. Note that the beam displacers used in our experiments always shift the photons with the horizontal polarization upward and keep the photons with the vertical polarization on the original path. Therefore, after the first two BDs, operation $U_1$ has been fulfilled. The beam splitter (BS) then gives two paths of different lengths $\ell_1$ and $\ell_2$. 
On the shorter path, an operation $U$ such that $U_2=UU_1$ is performed, while the longer path involves direct reflection to the measurement device. The ratio between these two paths is adjusted using a movable attenuator. The two HWPs after the first BS are fixed at angles of $-45^\circ$ and $45^\circ$, respectively.

In Bob's part, two operations are also enacted on the photons with probabilities $1-p$ and $p$ and the lengths of 
the two paths are also $\ell_1$ and $\ell_2$. The operation 
$V_1: (|0\rangle, |1\rangle) \rightarrow (|00\rangle, |10\rangle)$ is performed on the longer path and the operation 
$V_2: (|0\rangle, |1\rangle) \rightarrow (|01\rangle, |11\rangle)$ is performed on the shorter path. 
The photon B states $|01\rangle$, 
$|10\rangle$, $|00\rangle$, and $|11\rangle$ correspond, respectively, to the states $|\alpha \rangle$, $|\beta \rangle$, 
$|\gamma \rangle$, and $|\delta \rangle$ in Eq. \eqref{expst}. Unlike $U_1$ and $U_2$, the operations $V_1$ and $V_2$ can be completed using only BSs. The ratio between the two operations 
can also be tuned via a movable attenuator.

Here we use the time-bin method, 
that is, the coincidence detection only records the photon pairs who arrive the coincidence counter within the time window, to prepare the classical-classical states. As described above, when both photons of a pair travel the longer (shorter) paths, the operation $U_1~(U_2)$ is performed 
on photon $A$, and the operation $V_1~(V_2)$ is performed on 
photon $B$. Therefore, the state of the photon pair becomes 
$\rho = (1-p)(U_1\otimes V_1)\rho_d (U^\dag_1\otimes V^\dag_1) 
+ p(U_2\otimes V_2)\rho_d (U^\dag_2\otimes V^\dag_2)$, 
that is the same as Eq. (\ref{expst}). It should be noted that the time-bin method has been employed for coherently synthesizing quantum entangled states extensively in many previous works. However, here we can use it for incoherently producing classical correlations with suitable postselection.

After state preparation, determination of the method required to measure the prepared states is also a crucial problem. 
They are four-qubit states encoded in polarization and path degrees of freedom \cite{ZHDBCZLGZ,chaozhang2019, yunfenghuang2004, ojimenez2012}. As shown in Fig.({\ref{setup}}), the experimental setup contains four standard polarization tomography setups (SPTSs) that each consist of one half-wave plate (HWP), one quarter-wave plate (QWP), and a polarizing beam splitter (PBS). The first SPTS, on each side, is used to measure the polarization states and 
collapses these states to $|H\rangle$ so that the polarization information is erased. Then, the final BDs, on both sides,
convert the path information into polarization information to help the subsequent SPTSs to measure the path states. 
Using the four SPTSs, full quantum state tomography can be performed and the complete $16\times16$ density matrix $\rho$ can be reconstructed \cite{DanielF.V.James2001}. The reduced density matrices $\rho_A$ and $\rho_B$ are then derived from the experimentally determined 
state $\rho$, and the mutual information \eqref{mi} and 
the concurrence of the state $\rho_A$ are calculated.

\begin{figure}[!t]
 
\includegraphics[scale=0.325]{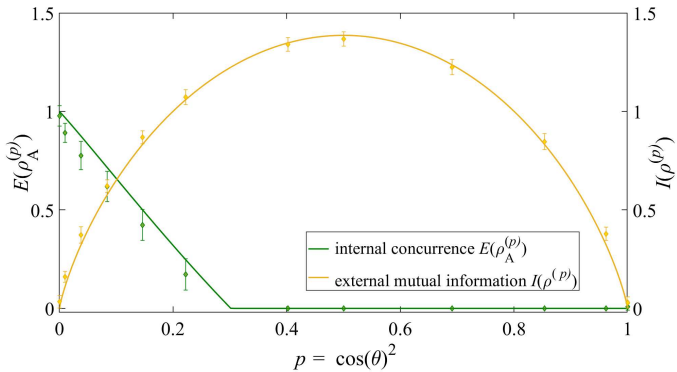}
\caption{\textbf{Experimental data.} The two curves present the theoretical values of the internal concurrence $E(\rho^{(p)}_A)$ (green) and the external mutual information $I(\rho^{(p)})$ (yellow), given by Eq. \eqref{Eexp} and Eq. \eqref{Iexp}, respectively, as functions of $p=\cos^2(\theta)$. The mutual information reaches a maximum when $p=0.5$. The concurrence $E(\rho^{(p)}_A)$ vanishes for $p > 0.302$. When $p=0$, the mutual information vanishes and the concurrence reaches its maximum value of $1$. When $p=1$, both the concurrence and the mutual information are zero. The experimental values (dots) correspond to 
states (\ref{expst}) with $\theta$ equal to $\{0$, $\pi/16$, $\pi/8$, $3\pi/16$, $\pi/4$, $9\pi/32$, $11\pi/32$, $3\pi/8$, $13\pi/32$, $7\pi/16$,  $15\pi/32$, and $\pi/2\}$. The error bars result from the quality of the interferometer and the fluctuations in the photon count.}\label{Ic}
\end{figure}

By adjusting the angle $\theta$ of the HWP before the BBO crystal, and the transmissivities of Alice's and Bob's 
attenuators so that $p=\text{cos}^2(\theta)$, 
we prepared 12 states. We used the values 
$0$, $\pi/16$, $\pi/8$, $3\pi/16$, $\pi/4$, $9\pi/32$, $11\pi/32$, $3\pi/8$, $13\pi/32$, $7\pi/16$, $15\pi/32$, and $\pi/2$ for $\theta$. The average fidelity of these states, as described using Eq. (\ref{expst}), is beyond $95\%$, see the 
Supplementary Materials \cite{fidelitysquare,SM}.
Fig. \ref{Ic} shows the theoretical and experimental values of the internal concurrence and external mutual information as functions of $p$.
%As Fig. \ref{Ic} shows, the concurrence, i.e., the internal quantum %nonseparability, decreases when $p$ increases and $\theta$ %decreases. The external correlation, which is quantified by the %mutual information, also varies with $p$, as Fig. \ref{Ic} shows. 
The experimental results for $\theta$ from $\pi/4$ to 
$\pi/2$, are represented as red dots with error bars in Fig. \ref{theoreticalexper}. They are all in the allowed region 
determined by inequality \eqref{mineq} with the function $\zeta$ given by Eq. \eqref{IC}, which experimentally demonstrates the 
tradeoff relation between internal nonseparability and external 
purely classical correlations.
The errors in the experiment mainly resulted from the quality of the interferometer, for which the visibility was approximately 50:1, and fluctuations in the photon count.

%%%%%%%%%%%%%%%%%%%%%%%%%%%%%%%%%%%%%%%%%%%%%%%%%%%%%%%%%%%%%%%%%%%%%%%%%%%%%%%%%%%%%%%%%%%%%%%%%%%%%%%%%%%%%%%%%%%%%%%%%%

\hfill

\noindent {\bf DISCUSSION} 

\noindent The entanglement and classical correlation tradeoff relation observed in the present experiment may be useful in 
the exploration of a number of fields, ranging from quantum communication \cite{nicolas2007} and quantum computation \cite{david1995} to open systems and many-body physics \cite{luigi2008}.
This relation does not apply solely to 
the considered optical system; 
it can also be observed in several other physical systems, including cold atoms and trapped ions. This tradeoff relation is thus a fundamental result for the development of quantum information science, particularly for quantum communication network.

For open systems \cite{opensystem}, external correlations 
between the system and the environment is a major concern because 
they may damage the entanglement inside the system. According to our results, external total correlations and internal quantum nonseparability limit each other. 
Even purely classical external correlations can have a detrimental effect on the internal quantum nonseparability. Therefore, to preserve the internal entanglement or nonseparability of the system as much as possible, the correlations, including the classical ones, between the system and the environment, must be reduced as low as possible.

In conclusion, we have presented the tradeoff relation between 
internal quantum nonseparability and external classical correlations in a photonic system experimentally. It is remarkable that the realization involves use of the time-bin method to produce purely classical correlations between 
the photons of a photon pair. Furthermore, polarization 
and path degrees of freedom of one of the photons of a pair 
have been entangled to realize the internal nonseparability experimentally. The classical-classical states are of major significance in this work and we have proposed a convenient and efficient method to prepare these states.

%\begin{figure}
%\includegraphics[scale=0.4]{I_p_2.eps}
%\caption{\textbf{Experimental values of external mutual information.} The external mutual information varies with $p=\text{cos}^2(\theta)$ smoothly. When $p=0$, only one term in Eq.(\ref{expst}) remains, where external correlation vanishes and the internal entanglement of system A reaches its maximal value. In the case of $p=1$, another term remains, where the state of system A is separable but also no external correlation is between system A and B.}\label{I_p}
%\end{figure}

%\begin{figure}
%\includegraphics[scale=0.3]{I_c_phi_exp.pdf}
%\caption{\textbf{Experimental values.} The experimental values correspond to different states as Eq. (\ref{expst}) shows, where the parameter $\theta$ is $\{0, \pi/16, \pi/8, 3\pi/16, \pi/4, 9\pi/32, 11\pi/32, 3\pi/8, 13\pi/32, \\7\pi/16, 15\pi/32, \pi/2\}$. Therefore the values of horizontal ordinate is not as uniform as Fig. \ref{Ictheory}. The errorbars come from the quality of interferometer and the photon count fluctutaions. }\label{Icexp}
%\end{figure}

%%%%%%%%%%%%%%%%%%%%%%%%%%%%%%%%%%%%%%%%%%%%%%%%%%%%%%%%%%%%%%%%%%%%%%%%%%%%%%%%%%%%%%%%%%%%%%%%%%%%%%%%%%%%%%%%%%%%%%%%%%

\hfill

\noindent {\bf DATA AVAILABILITY}

\noindent The data that support the findings of this study are available from the corresponding author upon reasonable
request.

%%%%%%%%%%%%%%%%%%%%%%%%%%%%%%%%%%%%%%%%%%%%%%%%%%%%%%%%%%%%%%%%%%%%%%%%%%%%%%%%%%%%%%%%%%%%%%%%%%%%%%%%%%%%%%%%%%%%%%%%%%

\hfill

\noindent{\bf ACKNOWLEDGEMENTS}

\noindent We thank Qiongyi He for helpful discussions. This work is funded by the National Natural Science Foundation of China (Grants Nos.~ 11674306, 92065113 and 11734015), National Key R$\&$D Program (No. 2016YFA0301300 and No. 2016A0301700), Anhui Initiative in Quantum Information Technologies, and the K.C. Wong Magna Fund in Ningbo University.

%%Chengjie Zhang :11734015

%%%%%%%%%%%%%%%%%%%%%%%%%%%%%%%%%%%%%%%%%%%%%%%%%%%%%%%%%%%%%%%%%%%%%%%%%%%%%%%%%%%%%%%%%%%%%%%%%%%%%%%%%%%%%%%%%%%%%%%%%%

\hfill

\noindent{\bf AUTHOR CONTRIBUTIONS}

\noindent S. Camalet, C.-J. Zhang and Y.-S. Zhang start the project and design the experiment. J. Zhu perform the experiment and complete the data analysis. S. Camalet and C.-J. Zhang provide the theoretical calculations. Y.-S. Zhang, C.-F. Li and G.-C. Guo supervise the project. J. Zhu, Y.-S. Zhang, C.-J. Zhang and S. Camalet write the manuscript and all authors participate in discussions. 

%\bibliographystyle{apsrev4-1}
%\bibliography{Bell}

%%%%%%%%%%%%%%%%%%%%%%%%%%%%%%%%%%%%%%%%%%%%%%%%%%%%%%%%%%%%%%%%%%%%%%%%%%%%%%%%%%%%%%%%%%%%%%%%%%%%%%%%%%%%%%%%%%%%%%%%%%

\hfill

\noindent{\bf ADDITIONAL INFORMATION}

\noindent The authors declare that they have no competing interests.

%%%%%%%%%%%%%%%%%%%%%%%%%%%%%%%%%%%%%%%%%%%%%%%%%%%%%%%%%%%%%%%%%%%%%%%%%%%%%%%%%%%%%%%%%%%%%%%%%%%%%%%%%%%%%%%%%%%%%%%%%%

%\hfill

%\begin{widetext}
%	
%\appendix
%\end{widetext}

\onecolumngrid
\setcounter{page}{1}
\renewcommand{\thepage}{Supplementary Material --\arabic{page}/1}
\setcounter{equation}{0}
\setcounter{figure}{0}
\setcounter{table}{3}
\renewcommand{\theequation}{S\arabic{equation}}

%
%\newpage
%\onecolumngrid
%\setcounter{page}{1}
\renewcommand{\thepage}{Supplementary Material --\arabic{page}/3}
\setcounter{equation}{0}
\setcounter{figure}{0}
\renewcommand{\theequation}{S\arabic{equation}}
\renewcommand{\thefigure}{S\arabic{figure}}

\section{Supplementary Material}

\section{Derivation of equation (4)}

When $A$ consists of two two-level systems, $E$ is the concurrence 
and $C$ is the mutual information, the function $\zeta$ mentioned 
in the main text is given by
$$ \zeta(c)=\max_{\boldsymbol \lambda \in \Lambda 
|h(\boldsymbol \lambda)=c} \max \Big\{ 0, 
\lambda_1-\lambda_3 - 2\sqrt{\lambda_2 \lambda_4} \Big\} , $$
where $\Lambda$ refers to the set of tuples of four probabilities 
summing to unity, arranged in decreasing order, and $h$ is the Shannon 
entropy, i.e., $h(\boldsymbol \lambda)=-\sum_i \lambda_i \ln \lambda_i$ \cite{C,VAD}.

To determine $\zeta$, we first consider $\chi$ defined by 
$$\chi(e)=\max_{\boldsymbol \lambda \in \Lambda 
|k(\boldsymbol \lambda)=e} h(\boldsymbol \lambda) , $$ where 
$k(\boldsymbol \lambda)
=\lambda_1-\lambda_3 - 2\sqrt{\lambda_2 \lambda_4}$. To find 
the maximum value of $h(\boldsymbol \lambda)$ under the constraints 
$k(\boldsymbol \lambda)=e$, $\sum_i \lambda_i=1$, and 
$0 \le \lambda_4 \le \lambda_3 \le \lambda_2 \le \lambda_1 \le 1$, 
we introduce $x$ and $y$ such that $0 \le x \le y \le 1$ and let 
$\lambda_4=x^2$ and $\lambda_2=y^2$. The conditions 
$k(\boldsymbol \lambda)=e$ 
and $\sum_i \lambda_i=1$ can be rewritten as $\lambda_{2 \mp 1}
=[1\pm e -(x \mp y)^2]/2$, and the above inequality requirements 
on the probabilities $\lambda_i$ determine the $e$-dependent domain 
${\cal D}_e$ of $(x,y)$ which exists for $e \in [-1/2,1]$. More precisely, 
${\cal D}_e$ is given by $x \ge 0$, $y \le -x+(1-e-2x^2)^{1/2}$, 
$y \le x/3+(3+3e-2x^2)^{1/2}/3$, and $y \ge -x/3+(3-3e-2x^2)^{1/2}/3$.

The first and second order derivatives of $h$ with respect to $x$ and $y$ 
can be written as 
\begin{eqnarray}
\partial_z h &=& -4z\ln z +(z-\bar z) \ln \lambda_1 + (z+\bar z) \ln \lambda_3 , 
\nonumber \\ 
\partial^2_z h &=& -4 \ln z -(1+e)/\lambda_1-(1-e)/\lambda_3 
+\ln (\lambda_1 \lambda_3), 
\nonumber \\ 
\partial^2_{z \bar z} h &=& (1+e)/\lambda_1-(1-e)/\lambda_3 
+\ln (\lambda_3/\lambda_1),
\nonumber
\end{eqnarray} 
where $z=x$ or $y$, $\bar x=y$, and $\bar y=x$. The function $h$ has a 
critical point in the interior of ${\cal D}_e$ for some values of $e$, but it is 
not a maximum, as shown by the fact that the Hessian determinant, 
$\partial^2_x h \partial^2_y h-(\partial^2_{xy} h)^2$, is negative at this point. 
For $e>0$, the boundary of ${\cal D}_e$ contains a line segment on the $y$-axis.
On this line segment, the maximum value of $h$ is at $y=\kappa(e)^{1/2}$ and is 
equal to $\mu(1+e-\kappa(e))+\mu(1-e-\kappa(e))-\kappa(e)\ln \kappa(e)$. 
The functions $\kappa$ and $\mu$ are defined in the main text. 
On the boundary of ${\cal D}_e$ and for $x>0$, the maximum value of $h$ 
is at $x=y=\sqrt{(1-e)/6}$ and is equal to 
$\mu(1+e)+\mu(1-e)+(1-e)\ln (3)/2$, or is reached in the limit $x \rightarrow 0$. 
Consequently, $\chi$ is given by the right side of eq. (4) in the main text. 

Since $\chi$ is a continuous and strictly decreasing function on 
$[-1/2,1]$, $\chi(-1/2)=2\ln 2$ and $\chi(1)=0$, it has an inverse 
function $\chi^{-1}$ with domain $X=[0,2\ln 2]$. Consider any 
$c \in X$ and define $e_c=\chi^{-1}(c)$. As seen above, there is 
$\boldsymbol \lambda \in \Lambda$ such that 
$k(\boldsymbol \lambda)=e_c=\chi^{-1}(c)$ and 
$h(\boldsymbol \lambda)=\chi(e_c)=c$. Let now $\boldsymbol \lambda$ 
be any tuple of $\Lambda$ such that $h(\boldsymbol \lambda)=c$ and 
let $e=k(\boldsymbol \lambda)$. Assuming $e>e_c$ implies 
$\chi(e)<\chi(e_c)=c$, and so, by definition of $\chi(e)$, 
$h(\boldsymbol \lambda)<c$. As this last inequality cannot hold, 
one has necessarily $k(\boldsymbol \lambda) \le \chi^{-1}(c)$, and hence
$$\max_{\boldsymbol \lambda \in \Lambda |h(\boldsymbol \lambda)=c} 
k(\boldsymbol \lambda)=\chi^{-1}(c). $$ 
When $\chi^{-1}(c) \le 0$, i.e., for $c \in [\ln (2\sqrt{3}),2\ln 2]$, 
$k(\boldsymbol \lambda)$ is non positive for any 
$\boldsymbol \lambda \in \Lambda$ such that $h(\boldsymbol \lambda)=c$, 
and so $\zeta(c)=0$. When $\chi^{-1}(c) > 0$, $\zeta(c)$ is equal to 
$\chi^{-1}(c)$, as given by eq. (4) in the main text.

\section{Derivation of equations (6) and (7)}

Consider a system $A$ consisting of two two-level systems and 
a four-level system $B$ in a classical-classical state of the form
$$\rho=p(1-q) | 00 \rangle \langle 00| 
\otimes |\alpha \rangle \langle \alpha| 
+(1-p)q |+ \rangle \langle + | \otimes |\beta \rangle \langle \beta|
+pq |11\rangle \langle 11| \otimes |\gamma \rangle \langle \gamma|
+ (1-p)(1-q)|- \rangle \langle - | \otimes |\delta \rangle \langle \delta| , $$
where $p,q \in [0,1]$, $|\alpha \rangle$, $|\beta \rangle$, 
$|\gamma \rangle$ and $|\delta \rangle$ are orthonormal states, 
$| 00 \rangle=| 0 \rangle \otimes | 0 \rangle$, 
$| 11 \rangle=| 1 \rangle \otimes | 1 \rangle$ and 
$|\pm\rangle
=(|0 \rangle \otimes | 1\rangle\pm|1 \rangle \otimes |0\rangle)/\sqrt{2}$ 
with $|0\rangle$ and $|1 \rangle$ denoting orthonormal states of 
the subsystems of $A$. Since $S(\rho)=S(\rho_A)=S(\rho_B)$, 
the corresponding mutual information between $A$ and $B$ is
$$I(\rho)=-p\ln p -(1-p)\ln (1-p) -q\ln q-(1-q)\ln (1-q) , $$
which simplifies to eq. (6) in the main text for $q=1-p$. 
Note that it is invariant under the changes $p \leftrightarrow 1-p$ 
and $q \leftrightarrow 1-q$.

The concurrence $E$ between the two subsystems of $A$ is determined 
by the eigenvalues $\mu_i$ of 
$\rho_A \sigma \otimes \sigma \rho_A^* \sigma \otimes \sigma$ 
where $\sigma=-|1\rangle\langle 0| +|0\rangle\langle 1|$ and 
$\rho_A^*=\rho_A$ is the complex conjugate of $\rho_A$ written in 
the standard basis $\{ | 00 \rangle,| 01 \rangle,| 10 \rangle,| 11 \rangle \}$. 
It reads as 
$E(\rho_A)=\max \{ 0, 2\max_i \sqrt{\mu_i}-\sum_i \sqrt{\mu_i} \}$. 
The eigenvalues $\mu_i$ are $q^2(1-p)^2$, $(1-q)^2(1-p)^2$, and twice 
$p^2q(1-q)$. They are given by the same expressions with $q$ replaced by 
$\tilde q=\min \{ q, 1-q \}=(1-|2q-1|)/2$. Since the last one is doubly 
degenerate, $2\max_i \sqrt{\mu_i}-\sum_i \sqrt{\mu_i}$ can be positive 
only when $\max_i \sqrt{\mu_i}=(1- \tilde q)(1-p)$, and so
$$E(\rho_A)=\max \{ 0, (1-2\tilde q)(1-p)-2p\sqrt{\tilde q(1-\tilde q)} \} , $$
which simplifies to eq. (7) in the main text for $q=1-p$. Clearly, $E(\rho_A)$ is 
also  invariant under the change $q \leftrightarrow 1-q$.

\section{Fidelities of prepared states}

\setlength{\tabcolsep}{1.5mm}
\begin{table}[!h]
\renewcommand\arraystretch{1.5}
\caption{Fidelities of prepared states}  
\centering
\scalebox{1}{
\begin{tabular*}{15cm}{ccccccc}  
\hline  
\hline
$\theta$    &0                  & $\frac{1}{16}\pi$ & $\frac{1}{8}\pi$  &                     $\frac{3}{16}\pi$  & $\frac{1}{4}\pi$  & $\frac{9}{32}\pi$ \\  
\hline  
fidelities  &$98.81\pm0.11\%$  & $95.66\pm0.35\%$ & $95.25\pm0.24\%$ & $95.30\pm0.31\%$  & $95.27\pm0.33\%$ & $95.80\pm0.22\%$\\  
\hline
\hline
$\theta$    &$\frac{11}{32}\pi$  & $\frac{3}{8}\pi$   & $\frac{13}{32}\pi$ &
$\frac{7}{16}\pi$   & $\frac{15}{32}\pi$ & $\frac{1}{2}\pi$ \\  
\hline  
fidelities  &$95.01\pm0.48\%$  & $95.55\pm0.42\%$ & $94.70\pm0.28\%$ &
$94.68\pm0.37\%$  & $95.64\pm0.26\%$ & $97.20\pm0.15\%$ \\
\hline
\end{tabular*}  }
\end{table}

\end{document}